\documentclass[aps,pre,twocolumn,showpacs,floatfix,preprintnumbers,amsmath,amssymb]{revtex4-1}
\usepackage{epsfig}
\usepackage{graphicx}
\usepackage{dcolumn}
\usepackage{bm}

\begin{document}
\title{Extreme events and event size fluctuations in biased random walks on networks}
\author{Vimal Kishore}
\email{phy.vimal@gmail.com}
\affiliation{Physical Research Laboratory,Navrangpura, Ahmedabad - 380009, India}
\author{M. S. Santhanam}
\email{santh@iiserpune.ac.in}
\affiliation{Indian Institute of Science Education and Research, 
Pune - 411021, India} 

\author{R. E. Amritkar}
\email{amritkar@prl.res.in}
\affiliation{Physical Research Laboratory, Navrangpura, Ahmedabad - 380 009, India.}


\begin{abstract}

Random walk on discrete lattice models is important to understand various
types of transport processes. The extreme events,
defined as exceedences of the flux of walkers above a prescribed threshold, have been studied
recently in the context of complex networks. This was motivated by the occurrence of
rare events such as traffic jams, floods, and power black-outs which take place on networks.
In this work, we study extreme events in a generalized random walk model in which the
walk is preferentially biased by the network topology. The walkers preferentially
choose to hop toward the hubs or small degree nodes. In this setting, 
we show that extremely large fluctuations in event-sizes
are possible on small degree nodes when the walkers are biased toward the hubs.
In particular, we obtain the distribution of event-sizes on the network.
Further, the probability for the occurrence of extreme events on any node in the network
depends on its 'generalized strength', a measure of the ability of a node to attract walkers.
The 'generalized strength' is a function of the degree of the node
and that of its nearest neighbors. We obtain analytical and simulation
results for the probability of occurrence of extreme events on the nodes of a network
using a generalized
random walk model. The result reveals that the nodes with a larger value of 'generalized strength',
on average, display lower probability for the occurrence of extreme events compared
to the nodes with lower values of 'generalized strength'.

\end{abstract}
\pacs{89.75.Hc, 05.40.-a, 05.40.Fb}
\maketitle

\section{Introduction}
Extreme events are typically associated with disasters of some kind or other, e.g.,
droughts, cold wave, cyclones, earthquakes, wind gusts and economic recession.
When a relevant variable, such as the wind speed $w(t)$ recorded at time $t$
in the case of wind gusts, exceeds certain prescribed threshold $q$ due to its
inherent fluctuations, {\it i.e.}, $w(t) > q$, then it is taken to be an
extreme event. In particular,
it is important to note that the magnitude of tremor, wind speed,
temperature, economic growth etc. are scalar variables.
A large number of results, both theoretical and empirical, are known about the
statistics and dynamics of extreme events for such univariate, scalar variables \cite{eevent1}. One significant result due
to classical extreme value theory is
that, depending on the probability distribution function of the variable,
the distribution of block maxima, for the uncorrelated sequence of random variables, converges to
only one of the three possible forms, namely, Fr\'echet, Gumbel and Weibull distributions \cite{coles}.

In contrast to this scenario, extreme events can also take place on complex networks.
Consider, for instance, the most common experience of web surfers; a web server
not responding due to the heavy load of http requests. This is an extreme event
taking place on the network of world wide web.
For example, the popular social networking site
Twitter handled about 600 tweets per second in early 2010 \cite{tweet}. 
According to an industry estimate, the Google search engine received approximately
34000 search requests per second by the end of 2009 \cite{google}. For most
websites on the world wide web that are unprepared for such a large number
of http requests, these numbers would represent extreme events and could
potentially disrupt the service. The power black-out in the north eastern
United States in 2003 is also an example of extreme event on the power 
transmission grid network. The cascading failures shut down more than 508 power generating units at 265 power plants
 during the peak of this black-out\cite{blackout}. Grid locks in highways
is an example of extreme event on transportation network. From the point of view
of physics, all these events could be thought of as an emergent phenomena
arising due to flux on the networks and could be regarded as extreme events arising
primarily due to limited handling capacity of the node. Transport on the networks continues to be
widely studied but much less attention has been focused on it from the point of
view of extreme events.  Generally, when the flux (packets of information or power or
highway traffic, in the case of examples given above) exceeds the handling capacity,
it turns out to be an extreme event for the particular node on the network.
In the earlier works related to congestion and cascade on networks \cite{congest1,congest2,congest3,congest4,
congest5,congest6,congest7,congest8}, handling capacity
is a key ingredient that needs to be prescribed upfront.

However, extreme events happen not only because of the limited handling capacity
of the node on a network but also because of inherent fluctuations in the flux passing
through the node. These fluctuations in the flux passing through a node
could be so large as they breach
a prescribed threshold, in which case, we label the event as an extreme event for the node.
This definition of extreme event for a node on any network is similar in spirit to
that of the classical extreme value theory. Then, a relevant question is
how the connectivity of the network affects the probability for extreme event occurrence.
By modeling the transport as standard random walks on networks,
it was shown in Ref. \cite{vsa} that the probability for the occurrence of extreme events $P(k_i)$, arising due to
inherent fluctuations, depends only on the degree $k_i$ of the $i$-th node in question.
In this work, the threshold $q_i$ was chosen to be proportional to typical fluctuation
size on $i$-th node. Thus, the extreme events are identified after taking care of the
natural variability of the flux passing through the given node.
Further, it was shown that, on average $P(k)$ is higher for small
degree nodes than for hubs.
This is a surprising result because it implies that, within the framework of
random walk on networks, even though hubs attract large flux (compared to small degree nodes)
they are less prone to extreme events. Thus, in the context of a node on a connected
network, larger flux does not necessarily translate
into higher probabilities for extreme events. This feature is one possible signature
of connectivity, {\it i.e.}, the network setting on which the system operates.
In contrast, for a scalar time series $w(t)$ larger flux would imply higher
extreme event probabilities.

Random walk on complex networks is a useful fundamental model against which to
compare other transport processes. Most realistic transport phenomena on networks, such as the
flux of information packets passing through the network of routers or road traffic, do not
proceed by performing random walk. In order to model the flux in a more realistic way, it
is useful to generalize the standard random walk to a situation in which the flux
is either biased toward hubs or small degree nodes. For example, consider the case of
two remote airports which are not directly connected by flights. Typically, they
would be connected through a major hub on the airline network. This is one practical
scenario in which the traffic is biased toward the hubs. This happens in many
a network settings; railways tend to connect the hinterland with the hubs, 
phones connect to nearest hubs on the network. Motivated by these physical examples,
in this work, we model the transport process as random walks biased by the topology of
the network and study
the extreme event probabilities and event-size distributions. We show that biased
random walk leads to extreme fluctuations in the event sizes on the network. In the
subsequent sections, we discuss the topologically biased random walk model on a
network and obtain analytical results for the probability of occurrence of extreme events
on any node. We show that the analytical and simulation results are in good agreement.

\section{Biased random walk on networks}
\subsection{Stationary distribution}
\begin{figure}[t]
\includegraphics*[width=3.2in,angle=0]{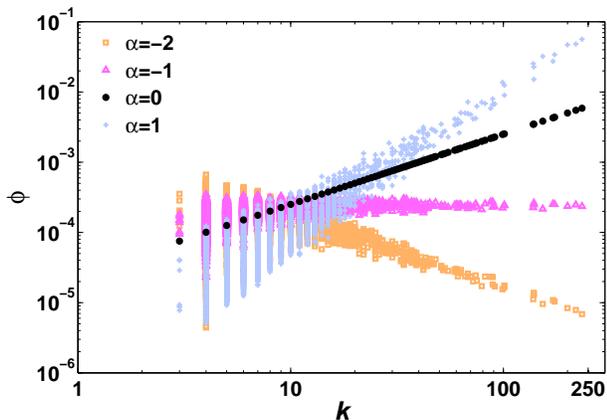}
\caption{(Color online) Strength $\phi$ as a function of degree $k$ for
different values of $\alpha$ in log-log plot.}
\label{k_vs_phi}
\end{figure}

We consider a connected, undirected, finite network with $N$ nodes and $E$ edges.
The network is characterized by a symmetric adjacency matrix ${\mathbf A}$ with
elements $A_{ij}=1$ if nodes $i$ and $j$ are connected by an edge and $A_{ij}=0$
otherwise. There are $W$ independent walkers performing biased random walk on
this network in the sense explained below. We denote by $b_{ij}$ the transition
probability for a walker to
hop from node $i$ to a neighboring node $j$. Let $P_{ij}$
be the probability that a walker starting at the node $i$ at time $n=0$ is at
node $j$ at time $n$. Then, the master equation can be written as
\begin{equation}
P_{ij}(n+1)=\sum_l A_{lj} ~ b_{lj} ~ P_{il}(n).
\label{master}
\end{equation}
The random walkers are biased by taking the time-independent transition probability
for hopping from $l$-th to $j$-th node to be \cite{fronczak1,fronczak2,yang}
\begin{equation}
b_{lj} \propto k_j^{\alpha},
\label{bias}
\end{equation}
where $\alpha$ is a parameter that defines the degree of bias imparted to the walkers.
Clearly, $\alpha=0$ corresponds to the standard random walk and the
transition probability is unbiased, where the walker can hop to any of the neighboring
node with equal probability.
For $\alpha > 0$, the random walkers are biased toward nodes
with larger degree or hubs. In contrast, if $\alpha < 0$, walkers preferentially
hop to small degree nodes. The larger (smaller) the $\alpha$, stronger the bias toward
the hubs (small degree nodes) is. Then, the normalized transition probability becomes
\begin{equation}
b_{lj}=\frac{k_j^{\alpha}}{\sum_{m=1}^{k_l} k_m^{\alpha}}.
\label{biasing}
\end{equation}
The summation in the denominator runs over the nearest neighbors of node $l$.
Using the transition probability in Eq.\ref{biasing}, the master equation becomes
\begin{equation}
P_{ij}(n+1)=\sum_l A_{lj}\frac{k_j^{\alpha}}{\sum_{m=1}^{k_l}k_m^{\alpha}}P_{il}(n).
\label{master1}
\end{equation}
By repeated iteration of Eq. \ref{master1}, it can be shown that $P_{ij}(n)$, as $n \to \infty$
leads to the stationary distribution
\begin{equation}
\lim_{n\to\infty} P_{ij}(n) = p_j = \frac{k_j^{\alpha}\sum_{l=1}^{k_j} k_l^{\alpha}}{\sum_{m=1}^{N} 
\left(k_m^{\alpha}\sum_{l=1}^{k_m} k_l^{\alpha}\right)}.
\label{statdist0}
\end{equation}
We can define the generalized strength of $j$ th node to be
\begin{equation}
\phi_j=k_j^{\alpha}{\sum_{i=1}^{k_j} k_i^{\alpha}},
\end{equation}
which is a measure of the ability of a node to attract walkers.
Note that $\phi_j$ depends on the bias parameter $\alpha$ and the degree of the
nearest neighbors to which it is connected by an edge. Hence, it is possible for the
nodes with same degree to have different generalized strengths. Thus, the generalized strength of the node is independent
of the global network structure but is dependent on the local connectivity
structure around the node. This is in contrast to the case of standard random walk (on networks)
in which large-scale structure of the network topology plays no significant role.
The local network structure is important for biased random walks on networks.
In Fig. \ref{k_vs_phi}, we show how the generalized strength $\phi$ depends on the degree of 
a node, for several values of $\alpha$, in a scale-free network
with degree exponent $\gamma=2.2$.
For $\alpha=1$ (crosses in Fig. \ref{k_vs_phi}), the generalized strength of a node is higher for large degree nodes (hubs) and an approximate linear relation is seen between $\phi_i$ and $k_i$
of $i$-th node. For $\alpha=0$, which is the standard random walk case, the generalized strength of the node is
the same as the degree of the node (solid circles in Fig. \ref{k_vs_phi}). However, for $\alpha=-1.0$,
$\phi$ is independent of $k$
especially for large degree nodes (triangles in Fig. \ref{k_vs_phi}).
In this case, the bias in the random walk represented by its generalized strength $\phi$ is balanced
by the degree of the node. In a scale-free network, a large number of small degree
nodes are present and they do not have identical values for the generalized strength $\phi$. This explains
the spread in $\phi$ for all values for $k < 50$.
Upon further decrease in the bias parameter $\alpha$ below -1.0 (open squares in Fig. \ref{k_vs_phi}),
nodes with a smaller degree or neighbors with smaller degree become important
and the generalized strength decreases with increasing degree.

\subsection{Extreme event probability}
The stationary distribution for the number of walkers in $j$-th node
can be rewritten in terms of the generalized strength $\phi$ as
\begin{equation}
p_j=\frac{\phi_j}{\sum_{l=1}^{N}\phi_l}.
\label{statdist}
\end{equation}
Thus, every node can be uniquely characterized by its generalized strength $\phi$. It is expected that two nodes
with the same value of $\phi$ show similar behavior as far as biased walks
on networks based on Eq. \ref{bias} are concerned.
In case of $\alpha=0$, we get $\phi_i=k_i$ and the stationary distribution simplifies
to $p_j=\frac{k_j}{2E}$, the result obtained for the case of standard random walk
in Ref. \cite{rw1}. Thus, in the case of a standard random walk, the degree $k$ characterises
the node. In the case of uncorrelated random networks,
the stationary occupation probability can be further simplified by using the mean field
approximation and can be written as \cite{fronczak1,fronczak2}
\begin{equation}
p_j = \frac{k_j^{\alpha+1}}{N\langle k^{\alpha+1}\rangle}.
\label{fronc}
\end{equation}
This approximate result suggests that the nodes with the same degree should have
the identical transition probabilities \cite{fronczak1}. This does not necessarily hold well
for the nodes of correlated networks such as the scale-free networks. This is because
in a scale-free network, the neighbourhood of nodes with identical degree are not
identical. Hence, to study extreme events we use Eq. \ref{statdist} instead of Eq. \ref{fronc}. 

\begin{figure}
\includegraphics*[width=3.2in,angle=0]{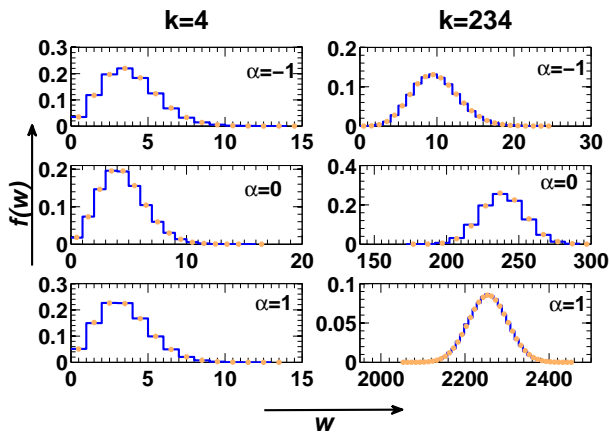}
\caption{(Color online) The distribution of walkers on two nodes with $k=4$ and $k=234$ for  ${\alpha} =-1.0,0.0$
and $1.0$. The solid lines show the distribution of walkers obtained from simulation while
solid circles belong to the binomial distribution obtained analytically using the stationary probability
in Eq. (\ref{statdist}).}
\label{bino_dist}
\end{figure}

Given that Eq. \ref{statdist} gives the probability to find one walker
on $i$-th node with generalized strength $\phi_i$, we can now obtain the distribution of
random walkers on $i$-th node. The formulation is applicable to any node on
the network and hence, in our further discussions, we suppress the index $i$ of the node.
Random walkers are independent and non-interacting and
hence the probability $f(w)$ of finding $w$ walkers on a node is $p^w$ while the rest of the walkers, $W-w$ are
distributed on the rest of the nodes of the network. When properly normalized, this leads to a
binomial distribution given by
\begin{equation}
f(w)={W \choose w}~p^w~(1-p)^{W-w}.
\label{binomial}
\end{equation}
The mean and variance of the flux passing through the given node is
\begin{eqnarray}
\langle f \rangle & = & W \frac{ \phi}{\sum_{l=1}^{N} \phi_l}, \nonumber \\  
\sigma^2 & = & W \frac{\phi}{\sum_{l=1}^{N} \phi_l} \left(1-\frac{\phi}{\sum_{l=1}^{N}\phi_l}\right).
\label{meanvar}
\end{eqnarray}
Note that the results in Eqs. \ref{binomial} and \ref{meanvar} depend only on the generalized strength $\phi$
that characterises a node including its neighbourhood. It does not depend on the large scale
connectivity pattern. Hence, these results will hold good for any network, such as scale-free, random
or small world, irrespective of its degree distribution. Further, in the cases for which
$\sum_{l=1}^{N}\phi_l >> \phi$, we obtain the approximate relation $\sigma \approx \langle f \rangle^{1/2}$.
This relation can be thought of as a generalization of a similar relation for the unbiased
random walks reported in Ref. \cite{vsa}. However, the exponent $1/2$ is not universal and instead depends
on details such as the fluctuation in number of walkers and sampling resolution of the flux \cite{breakdown}.
The distribution of random walkers on two nodes with different degrees, $k=4$ and $k=234$,
is plotted in Fig. \ref{bino_dist}. The biased random walk simulations were performed on a
scale-free network with 5000 nodes with 19915 links and 39830 walkers. Initially, at time $n=0$, the walkers
are randomly distributed on $N$ nodes. 
The simulation results presented in Fig. \ref{bino_dist} have been obtained after
averaging over 100 realisations with different initial conditions of random walkers. 
The simulation
results, the solid lines in Fig. \ref{bino_dist}, show a good agreement with the
analytical distribution given by Eq. \ref{binomial}. 

\section{Probability for extreme events}
\begin{figure}[t]
\includegraphics*[width=3.2in,angle=0]{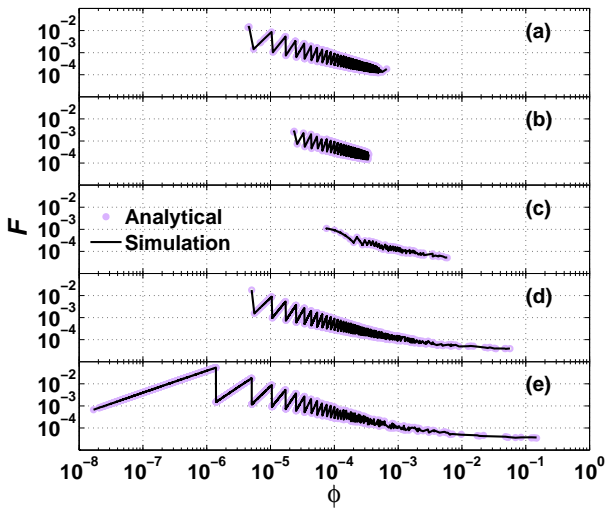}
\caption{(Color online) The probability for the occurrence of extreme events plotted 
as a function of node generalized strength $\phi$ (normalized) for different values of bias parameters 
(a) $\alpha=-2.0$, (b) $\alpha=-1.0$, (c) $\alpha=0.0$, (d) $\alpha=1.0$ and (e) $\alpha=2.0$. 
The threshold for extreme event is $q= \langle f \rangle +4\sigma$. The circles are from 
analytical results in Eq. (\ref{xvp}) while solid lines are the simulation results performed on a scale-free network
($N=5000$, $E=19915$) with $W=2E$ walkers averaged over 100 realizations with randomly chosen initial positions of walkers.}
\label{xvprob}
\end{figure}

We take an extreme event to be the one for which the probability of occurrence is
small and is typically associated with the tail of the probability distribution
function for the events. We extend this principle to the events on the nodes of
a network \cite{vsa}. Given that the number of walkers $w$ passing through a node with
generalized strength $\phi$ follow the Binomial distribution, if more than $q$ walkers pass through
the node, then it is an extreme event for the node. Then, the probability
for the occurrence of extreme event is
\begin{eqnarray}
F_i & = & \sum_{w=q_i}^W   ~ {W \choose w} ~{p}_i^w ~ (1-{p}_i)^{W-w}, \label{xvp0} \\
              & = & I_{p_i}(\lfloor q_i \rfloor + 1, W - \lfloor q_i \rfloor ),
\label{xvp}
\end{eqnarray}
where $\lfloor u \rfloor$ is the floor function defined as the largest integer not greater than $u$
 and $I_z(a,b)$ is the standard incomplete Beta function \cite{lopez}. In this form,
the extreme event probability will depend on the choice of threshold $q_i$.
First, we consider the case of constant threshold.
If $q_i=0$, using Eq. \ref{xvp0} we obtain $F_i = 1$
for all the nodes on the network. Thus,
all the nodes will experience extreme events all the time.
On the other hand, if we set $q_i=W$, then we obtain 
\begin{equation}
F_i = p_i^W.
\end{equation}
Since $p_i << 1$, we get $F_i \approx 0$ for
all the nodes implying that there are no extreme events anywhere
in the network. Hence, these choices of threshold values are not 
physically interesting cases. Any other arbitrary choice such as
$q_i=q_0$, where $q_0$ is a constant, will predominantly lead to some nodes
encountering extreme events nearly all the time and others
having no events at all. This too is not an interesting case.
The foregoing arguments imply that an interesting scenario would
arise if the threshold is chosen to be proportional to
the natural variability of the flux passing through a node.
Thus, we choose the threshold for extreme events to be \cite{vsa}
\begin{equation}
q_i = \langle f_i \rangle + m \sigma_i,
\end{equation}
where $m \ge 0$. The mean flux $\langle f_i \rangle$ and
standard deviation $\sigma_i$ are given by Eq. \ref{meanvar}.
Substituting $q_i$ in Eq. \ref{xvp}, it is clear
that the probability for the occurrence of extreme events
is dependent only on the generalized strength $\phi$ of the node. In Fig. \ref{xvprob},
we show the simulation and analytical results for the probability
of extreme events as a function of $\phi$ for several choices of $\alpha$.
The numerical results are based on simulations with $W=39380$ walkers
on a scale-free network with $N=5000$ nodes evolved for $10^7$ time steps.
An unusual feature is that $F_i$ predicts higher probability of occurrence
of extreme events, on average, for nodes with small values of generalized strength $\phi$ than for
the nodes with higher values of generalized strength $\phi$. For instance, in Fig. \ref{xvprob}(a),
the probability of extreme event occurrence is generally higher for nodes with $\phi < 10^{-5}$
than for nodes with $\phi > 10^{-3}$. A similar effect is seen in Figs. \ref{xvprob}(b)- \ref{xvprob} (e).
Even though nodes with higher the generalized strength $\phi$
attract more walkers as given by Eq. \ref{statdist0}, this does not imply that they
also have higher probability for extreme events.
This is a generalization of the
result obtained in Ref. \cite{vsa} for the standard random walk on networks 
which shows that the extreme events are
more probable for nodes with small degree than for the ones with high degree.
The local fluctuations seen in Fig. \ref{xvprob} are inherent in the system
and not due to insufficient ensemble averaging.  Further, notice that
Eq. \ref{xvp} does not depend on the large scale structure of the topology and
hence it is valid for biased random walks on any topology, random
or small-world or scale-free.

However, the local connectivity patterns in the vicinity of any node plays
a crucial role in the diffusion of an extreme event. Suppose an extreme event
takes place at node $A$ at time $n$, then one interesting question is how
probable it is for an extreme event to take place in its immediate
neighborhood at time $n+1$, i.e, after the first jump. We call it first-jump probability and it is similar to the one reported in \cite{arw1}.
In the case of a
standard random walk ($\alpha = 0$), our simulations (not shown here) indicate that in general
if node $A$ is a hub, then the probability to encounter an extreme event
in its neighbourhood is higher (at least by a factor of 3-4) compared to the case when node $A$ is a
small degree node. For biased random walks, the results suggest a higher
likelihood for an extreme event to be transferred to its neighbourhood in the case
when $\alpha < 0$ compared to the case with $\alpha > 0$.

\section{Fluctuations in event size}
The size of an event is measured in units of the standard deviation $\sigma$ of the
flux passing through a node. In this section, we show that the extreme fluctuations
in the flux of walkers are realised in the case of $\alpha =  2$ which implies
that the walkers are biased toward the nodes with larger generalised strength $\phi$ (hubs).
An event is of size $m$ if $m\sigma \leq w - \langle w \rangle < (m+1)\sigma$,
where $w$ is the number of walkers on a given node.
 
\begin{figure}[t]
\centerline{\includegraphics*[width=3.5in,angle=0]{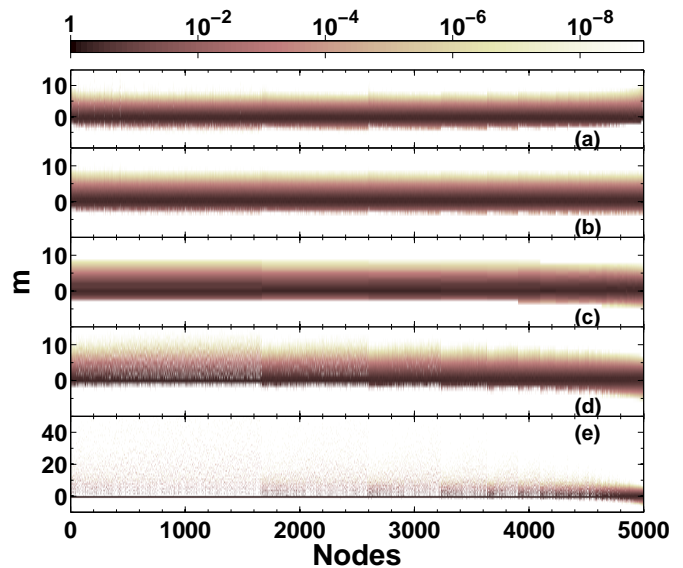}}
\caption{(Color online) The distribution of event sizes for biased random walks as a function
of node number on $x$-axis obtained from simulations performed on a scale-free network for different values of bias parameters 
(a) $\alpha=-2.0$, (b) $\alpha=-1.0$, (c) $\alpha=0.0$, (d) $\alpha=1.0$ and (e) $\alpha=2.0$. 
The nodes are arranged in the order of increasing degree. The probability values $\mathcal{P}_m$ are color coded. 
This should be compared with analytical results in Fig. \ref{anaesize}.}

\label{simesize}
\end{figure}

Then, the probability for the occurrence of an event of size $m$ can be written down as,
\begin{equation}
\mathcal{P}_m = I_p(\lfloor q_m \rfloor + 1, W - \lfloor q_m \rfloor ) -
                I_p(\lfloor q_{m+1} \rfloor + 1, W - \lfloor q_{m+1} \rfloor ).
\label{esizedist}
\end{equation}
To illustrate the result, we show the distribution of event sizes in Fig. \ref{simesize}
for $\alpha=-2,-1,0,1,2$ in a scale-free network obtained from simulations evolved for 
$10^7$ steps and averaged over $100$ ensembles.
Here, the events with probability of occurrence of less than $10^{-8}$ have been discarded to maintain the numerical accuracy.
In the case of $\alpha=0$ (standard random walk), the distribution of events is shown in Fig. \ref{simesize}(c).
The events of size $m=0$ are highly probable with $\mathcal{P}_0 \sim 0.1$. In contrast, the probability
for the events of size $|m| > 0$
decrease and thus the extreme events of size $m=-2,8$ occur with probability
$\mathcal{P}_{-2} \sim \mathcal{P}_8 \sim 10^{-8}$. The
limitation on the lower limit of event sizes is restricted
by the minimum possible number of walkers on a node, i.e., $0$. For lower degree nodes,
events of sizes $-2\sigma$ to $8\sigma$ are observed
but in the case of higher degree nodes $k>100$, events sizes range from $-5\sigma$ to $6\sigma$ only. In the case of a
standard random walk, for the whole network, event size $m$ varies from $-5\sigma$ to $8\sigma$.

\begin{figure}
\centerline{\includegraphics*[width=3.5in,angle=0]{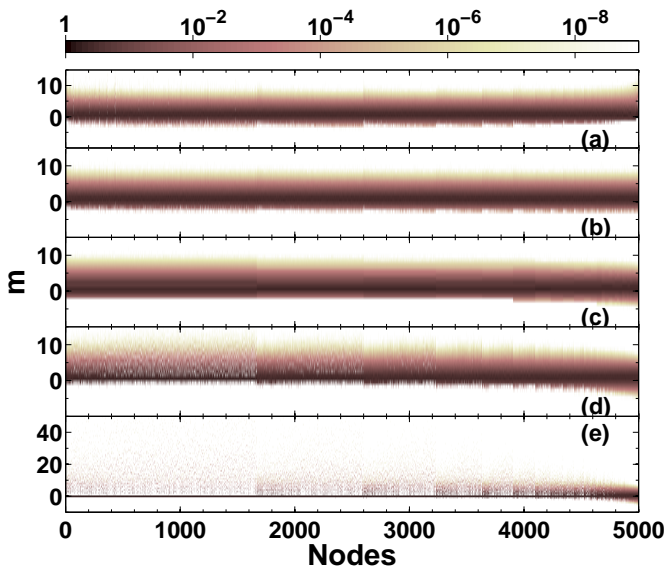}}

\caption{(Color online) The distribution of event sizes for biased random walks as a function
of node number on $x$-axis obtained analytically using Eq. \ref{esizedist} for different values of bias parameter (a) $\alpha=-2.0$, 
(b) $\alpha=-1.0$, (c) $\alpha=0.0$, (d) $\alpha=1.0$ and (e) $\alpha=2.0$. 
The nodes are arranged in the order of increasing degree. The probability values $\mathcal{P}_m$ are color coded.}
\label{anaesize}
\end{figure}

In comparison, for the case of $\alpha=1$ shown in Fig. \ref{simesize}(d) the events of size $8$ have higher probability 
of occurrence ($\mathcal{P}_8\sim10^{-7})$ and events of even higher sizes are also possible. 
For $\alpha=2$, even higher size events, as large as 40, become highly probable
for small degree nodes as seen in Fig. \ref{simesize}(e). Thus, in general, 
for larger $\alpha$, larger size events become probable when compared
with the case of $\alpha=0$. Physically, this can be understood as follows. With
$\alpha=0$, the random walkers perform unbiased random walk. However,
for $\alpha=2$, the walkers preferentially choose to hop to nodes
with larger degree (hubs). Since large degree nodes are mostly well connected
among themselves, very few walkers reach small degree nodes. Hence the
average flux through the small degree nodes becomes so small that even
occasional visits by a few walkers lead to extremely large size events.
These occasional visits lead to probability of order $10^{-6}$ even
for events of size 40. Hence, in the case of biased random walks, 
extremely large fluctuations in event sizes can be observed in small
degree nodes. This effect is also seen in the analytical results obtained using Eq. \ref{esizedist} shown in
Fig. \ref{anaesize}. 

On the other hand, for cases $\alpha=-2,-1$ such large fluctuations are not visible in 
the event sizes in Fig. \ref{simesize}(a) and \ref{simesize}(b).  For $\alpha=-1$ in Fig. \ref{simesize}(b),
there is a small increase in the event sizes (when compared to $\alpha=0$) for the small degree
nodes but it is not as large as in $\alpha=1$ case. Further, with  $\alpha=-1$, it must
also be noted that the probability profile remains similar for most of the nodes irrespective of
the large differences in their degree. This is because $\phi$ is an approximate constant for
most of the nodes since, in this case, the effect of the bias is balanced by the degree of these nodes.
For $\alpha=-2$, the flux is strongly biased towards small degree nodes and again events of
sizes $m=10$ can be seen in Fig. \ref{simesize}(a) though only on the higher degree nodes.
The event sizes for hubs are not as large as observed in case of $\alpha=2$ for lower degree nodes.
It can be explained as follows; when $\alpha=-2$, the flux preferentially flows through
the small degree nodes which form the bulk in a scale-free network. Most small degree nodes do not
have a direct link with other small degree nodes but are connected
through a hub. Hence, despite the biased walk favoring the small degree nodes, sufficiently large flux
flows through the hubs as well. Hence, abnormally large event size fluctuations are not seen
in hubs for $\alpha=-1, -2$. All these features show a good agreement with the analytical result
obtained in Eq. \ref{esizedist} and shown in Fig. \ref{anaesize}.

\section{Discussion and summary}

This work is an attempt to understand the extreme events occurring on the nodes
due to flow on networks which typically is directed toward or away from the hubs. In this work, 
we study a biased random walk model in which the traffic preferentially moves either toward
or away from the hubs and we analytically obtain the probabilities for the occurrence
of extreme events. In this framework, extreme events are due to inherent fluctuations
in the flux passing through any node and is defined as exceedences above a chosen
threshold $q$. The threshold is chosen to be proportional to the natural variability
of the node.
Each node on the network is characterized by generalized strength $\phi$ which depends
on its degree and that of its immediate neighbourhood. It is a measure of
how much traffic is attracted to the particular node. The larger the generalized strength of
a node is, larger its ability to attract walkers.
In this paper, we have shown that the nodes with a smaller generalized strength, on an average,
have a higher probability for the
occurrence of extreme events when compared to nodes with higher generalized strength. Further,
we have also shown that when the flux is biased toward the hubs, abnormally
large fluctuations in event sizes become highly probable. This is one
possible signature of the topologically biased flow in a scale-free network.

In general, it is possible to conceive of many ways by which bias can be imparted
to independent random walkers on networks. These biasing strategies are motivated by 
real observations and the quest for efficient search strategies on networks. Various kind of biases based on the local environment, 
shortest paths, the entropy of random walk and various adaptive strategies are some examples
of biased random walk on networks \cite{brw,shw,merw1,merw2,arw1,arw2}.
It will be interesting to study the extreme event
probabilities under such biasing strategies. However, we emphasise that if the stationary
probability distribution equivalent to Eq. \ref{statdist0} exists for all the above strategies, then
it would be possible to define extreme events and analyze them following
the methods presented in this work.

In the context of scale-free network, it has been argued that hubs are important for better functioning of the network. 
Apart from being responsible for providing better connectivity, existence of hubs makes the scale-free network robust 
against the random node removal but fragile if the node removal is targeted \cite{node_del1,node_del2}. 
The results in this paper show that extreme events due to natural fluctuations are more probable on small degree nodes
(when compared to the hubs). Hence special attention must be paid
to designing the capacity of the small degree nodes so that extreme
events can be smoothly handled without leading to disruption of the node.
The results in this paper can be used to estimate the
capacity a node should possess if it should handle extreme events
of size, say, $m$. If we want the node to handle $4\sigma$ events
smoothly, then the required capacity can be obtained by inverting
Eq. \ref{xvp}. Thus, the numbers so obtained can be useful as an
input for arriving at a capacity to be built for the nodes on a network.

\begin{acknowledgements}
The simulations were carried out on computer clusters at PRL, Ahmedabad and IISER Pune.
VK would like to thank IISER Pune for the hospitality provided during his stay in Pune.
\end{acknowledgements}


\begin{thebibliography}{99}

\bibitem{eevent1} S. Albeverio, V. Jentsch and Holger Kantz (Ed.) , {\it Extreme events
in nature and society}, (Springer, 2005).

\bibitem{coles} Stuart Coles {\it An Introduction to Statistical Modeling of Extreme Values},(Springer, London, 2001)

\bibitem{tweet} \url{ http://blog.twitter.com/2010/02/measuring-tweets.html}

\bibitem{google} \url{ http://www.comscore.com/Press_Events/Press_Releases/2010/1/Global_Search_Market_Grows_46_Percent_in_2009}
\bibitem{blackout} U.S.-Canada Power System Outage Task force (April, 2004), {\it see} \url{https://reports.energy.gov}
\bibitem{congest1} D. De Martino, Luca Dall'Asta, Ginestra Bianconi and Matteo Marsili, Phys. Rev. E {\bf 79}, 015101(R) (2009).
\bibitem{congest2} P. Echenique, J. Gomez-Gardenes and Y. Moreno, EPL {\bf 71}, 325 (2005).
\bibitem{congest3} K. Kim, B. Kahng and D. Kim, EPL {\bf 86}, 58002 (2009).
\bibitem{congest4} B. Tadic, G. J. Rodgers and S. Thurner, Int. J. Bifurcation Chaos Appl. Sci. Eng.{\bf 17}, 2363 (2007). 
\bibitem{congest5} Douglas J. Ashton, T.C. Jarrett and N. F. Johnson,  Phys. Rev. Lett. {\bf 94}, 058701 (2005).
\bibitem{congest6} Liang Zhao, Ying-Cheng Lai, Kwangho Park and Nong Ye,  Phys. Rev. E {\bf 71}, 026125 (2005).
\bibitem{congest7} R. Germano and A.P.S. de Moura,  Phys. Rev. E {\bf 74}, 036117 (2006). 
\bibitem{congest8} W. Wang, Zhi-Xi Wu, R. Jiang, G Chen and Y C Lai Chaos {\bf 19}, 033106 (2009).

\bibitem{vsa} Vimal Kishore, M. S. Santhanam and R. E. Amritkar, Phys. Rev. Lett {\bf 106}, 188701 (2011).


\bibitem{fronczak1} Agata Fronczak and Piotr Fronczak, Phys. Rev. E {\bf 80}, 016107 (2009); 
\bibitem{fronczak2} Wen-Xu Wang, B.H. Wang, C. Y. Yin, Y.B. Xie and T. Zhou, Phys. Rev. E {\bf 73}, 026111 (2006).

\bibitem{yang} S.-J. Yang, Phys. Rev. E {\bf 71}, 016107 (2005).
\bibitem{rw1}  J. D. Noh and H. Rieger, Phys. Rev. Lett. {\bf 92}, 118701 (2004).
\bibitem{breakdown}  S. Meloni, J. Gomez-Gardenes, Vito Latora and Y. Moreno, Phys. Rev. Lett. {\bf 100}, 208701 (2008).
\bibitem{lopez} J. L. Lopez and J. Sesma, Integral Transforms and Special Functions {\bf 8}, 233 (1999).

\bibitem{brw} V. Sood and P. Grassberger, Phys. Rev. Lett {\bf 99}, 098701 (2007).
\bibitem{shw} Shengyong Chen, Wei Huang, Carlo Cattani, and Giuseppe Altieri, Mathematical Problems in Engineering, {\bf 2012}, 732698 (2012).
\bibitem{merw1} Z. Burda, J. Duda, J. M. Luck, and B. Waclaw, Phys. Rev. Lett. {\bf 102}, 160602 (2009)
\bibitem{merw2}R. Sinatra, J. Gómez-Gardeñes, R. Lambiotte, V. Nicosia, and V. Latora, Phys. Rev. E {\bf 83}, 030103(R) (2011).
\bibitem{arw1} B. Tadic, Eur. Phys. J. B {\bf 23},221 (2001).
\bibitem{arw2} A.N. Mian, R. Baldoni, R. Beraldi, {\it Proceedings of the 29th IEEE International Conference on Distributed Computing Systems Workshops, Montreal, QC} (IEEE, Piscataway, NJ,2009). pp. 153-157.

\bibitem{node_del1} R. Cohen, K. Erez, ben-Avraham D. and S. Havlin, Phys. Rev. Lett {\bf 85}, 4626 (2000).
\bibitem{node_del2} R. Cohen, K. Erez, ben-Avraham D. and S. Havlin, Phys. Rev. Lett {\bf 86}, 3682 (2001).






\end{thebibliography}
\end{document}